\documentclass{webofc}
\usepackage[varg]{txfonts}   
\begin{document}
\title{400Gbps benchmark of XRootD HTTP-TPC}
%
%

\author{\firstname{Aashay} \lastname{Arora}\inst{1}\fnsep\thanks{\email{aaarora@ucsd.edu}} \and
        \firstname{Jonathan} \lastname{Guiang}\inst{1}\fnsep\thanks{\email{jguiang@ucsd.edu}} \and
        \firstname{Diego} \lastname{Davila}\inst{1}\fnsep\thanks{\email{didavila@ucsd.edu}} \and
        \firstname{Frank} \lastname{Würthwein}\inst{1}\fnsep\thanks{\email{fkw@ucsd.edu}} \and
        \firstname{Justas} \lastname{Balcas}\inst{2}\fnsep\thanks{\email{jbalcas@caltech.edu}} \and
        \firstname{Harvey} \lastname{Newman}\inst{2}\fnsep\thanks{\email{newman@hep.caltech.edu}}
}

\institute{University of California, San Diego, La Jolla, CA, USA \and California Institute of Technology, Pasadena, CA, USA}

\abstract{Due to the increased demand of network traffic expected during the HL-LHC era, the T2 sites in the USA will be required to have 400Gbps of available bandwidth to their storage solution. 
With the above in mind we are pursuing a scale test of XRootD software when used to perform Third Party Copy transfers using the HTTP protocol. Our main objective is to understand the possible limitations in the software stack to achieve the target transfer rate; to that end we have set up a testbed of multiple XRootD servers in both UCSD and Caltech which are connected through a dedicated link capable of 400 Gbps end-to-end. Building upon our experience deploying containerized XRootD servers, we use Kubernetes to easily deploy and test different configurations of our testbed. In this work, we will present our experience doing these tests and the lessons learned.}
\maketitle
\section{Introduction} 
\label{sec:introduction}
The High Luminosity LHC data collection era, planned to start in 2027, poses hard network challenges for the Tier 2 Sites in the US. All sites will require 400 Gbps burst capabilities for hours and the steady state network bandwidth consumption is expected to be approximately 100 Gbps, depending on the operational details and use of the various event-formats used in high energy physics \cite{carder2022basic}. Given these requirements, in addition to making sure that the hardware is up to par, it becomes imperative that we verify the robustness of our software stack to make sure it can support this throughput. The use of the HTTPS protocol has been steadily increasing on the internet to encrypt and secure the communication between servers which makes it a strong contender to ensure fast encryption and high security. Based on the success of HTTPS as a protocol for third-party-copy (TPC) transfers showcased in a past study \cite{Fajardo2021}, the adoption to use HTTPS-TPC for high energy physics data transfers has already been completed. In conjunction with XRootD \cite{xrootd, xrootd-paper} as the storage backend, which provides high-performance, scalable fault-tolerant access to data repositories of many kinds, XRootD HTTPS-TPC transfers are presently the projected setup for exabyte scale transfers. 
In this paper, we test the feasibility of using this setup for the aforementioned high throughput requirements by deploying XRootD storage servers at the University of California, San Diego (UCSD) and the California Institute of Technology (Caltech) respectively and running TPCs between them. The specifics of the deployment are highlighted in Section \ref{sec:setup}. The results are presented in Section \ref{sec:result}, additionally a study to parameterize the throughput using the number of concurrent single stream 1~GB transfers (streams hereafter) and latency between the endpoints is presented in Section \ref{subsec:extended}.

\section{Previous Studies} 
\label{sec:previous}
In order to compare the performance of XRootD HTTPS-TPC with the GridFTP protocol ~\cite{gridftp} which was the standard for LHC transfers in the past, we deployed XRootD servers on several nodes in the National Research Platform \cite{nrp} (formerly Pacific Research Platform) Kubernetes \cite{k8s} Cluster connected to the wide area network (WAN) at 100 Gbps and performed TPCs among them \cite{Fajardo2021}. The results revealed that XRootD HTTPS-TPC performs slightly better than GridFTP on average over high throughput links.

\begin{figure}
    \centering
    \includegraphics[width=0.7\textwidth]{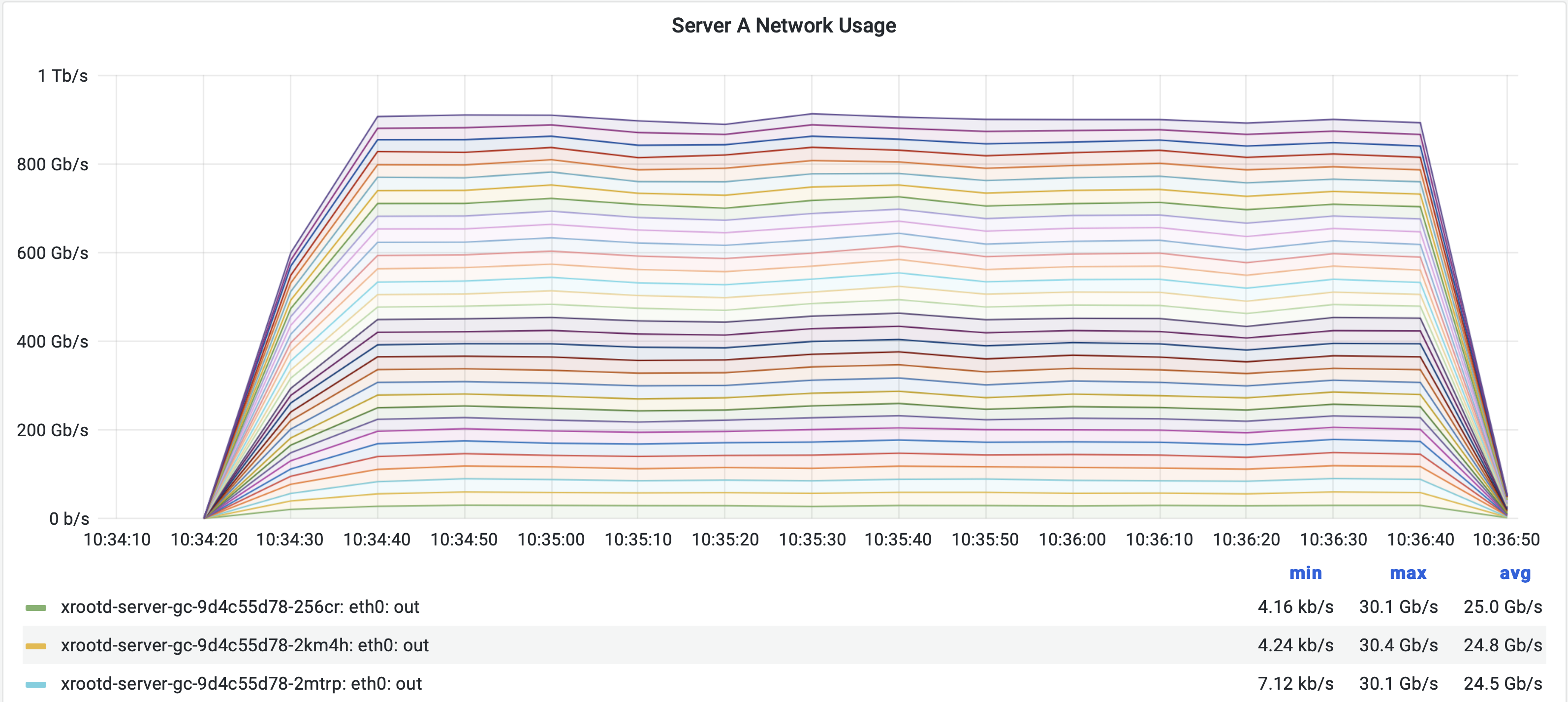}
    \caption{Monitoring Plot showing 1 Tbps flowing out of an XRootD cluster deployed in Microsoft Azure}
    \label{fig:azure}
\end{figure}

This result evoked the urge to test the limits of XRootD HTTPS over high-bandwidth links. A natural choice was the cloud, namely Microsoft Azure \cite{azure} in our case, which gave us access to dedicated ingresses and industry-standard processors and memory thus providing us with a high-performance testbed without disturbing production workflows. Repeating the throughput benchmarking exercise in this environment, we were able to get 1 Tbps \cite{arora} on low latency links between endpoints in the same cloud region (US-West) as showcased in Figure~\ref{fig:azure}. 

Building upon the insights gained from our previous tests, the next thing to do was to test this over WAN with 400 Gbps connectivity. We deployed XRootD servers on endpoints with appreciable latency between them and performed TPCs between them as detailed in the rest of this paper.

\section{Setup} 
\label{sec:setup}

\subsection{Deployment Details} 
\label{subsec:deployment}
Our testbed is comprised of two sets of servers, one at each of the participating institutions --- UCSD and Caltech. The two sets of servers are geolocated 120 miles apart on the west coast of the US, with a dedicated link of 400 Gbps and a network latency of roughly 3ms connecting both sites. This link is mostly dedicated as it is only shared with other R\&D projects.
Each group of servers is configured as an XRootD cluster i.e. all data servers connect to a redirector that acts as a load balancer.

\begin{figure}[ht!]
    \centering
    \includegraphics[width=\textwidth]{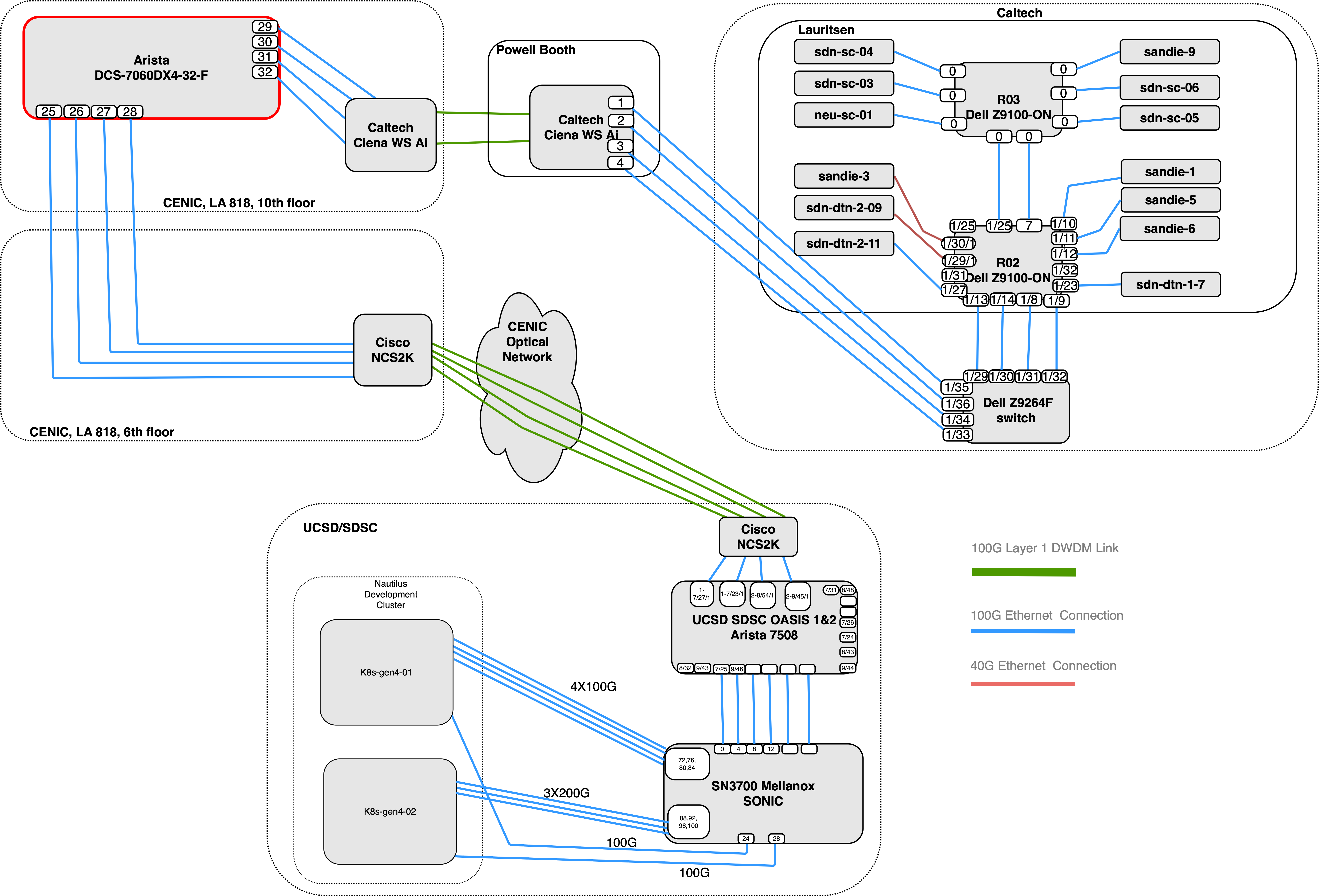}
    \caption{Network connectivity between UCSD and Caltech}
    \label{fig:400-diagram}
\end{figure}

At Caltech we have 13 servers, the specifications of which can be seen in Table~\ref{tb:specs-caltec}. At UCSD we have 2 servers with their specifications described in Table~\ref{tb:specs-ucsd}. 
In Figure~\ref{fig:400-diagram} we can see the diagram that depicts the network connectivity between both sites. It is evident that a minimum bandwidth of 400 Gbps is available between the two sites.

\begin{table}[htb!] 
\vspace{-0.2cm} 
\scriptsize
\centering
\caption{Caltech servers specifications}
\vspace{-0.3cm} 
\begin{tabular}{|r|r|r|r|r|} \hline
{\shortstack{Name}} & {\shortstack{CPU model}}  & {\shortstack{\# of real\\cores}} & {\shortstack{RAM\\GB}} & {\shortstack{Total bandwidth\\capacity Gbps}} \tabularnewline \hline \hline
sandie-1 & 2x E5-2667 v3 @ 3.20GHz & 16 & 256 & 100 \tabularnewline \hline
sandie-3 & 2x Silver 4110 CPU @ 2.10GHz & 16 & 96 & 40 \tabularnewline \hline
sandie-5 & 1x AMD 7551P @ 2Ghz & 32 & 256 & 100 \tabularnewline \hline
sandie-6 & 1x AMD 7551P @ 2Ghz & 32 & 256 & 100 \tabularnewline \hline
sdn-dtn-1-7 & 2x E5-2687W v3 @ 3.10GHz & 20 & 128 & 100 \tabularnewline \hline
sdn-dtn-2-09 & 2x E5-2690 v2 @ 3.00GHz & 20 & 128 & 40 \tabularnewline \hline
sdn-dtn-2-11 & 2x E5-2670 v3 @ 2.30GHz & 24 & 128 & 100 \tabularnewline \hline 
neu-sc-01 & 2x E5-2667 v4 @ 3.20GHz & 16 & 128 & 100 \tabularnewline \hline
sdn-sc-03 & 2x E5-2667 v4 @ 3.20GHz & 16 & 128 & 100 \tabularnewline \hline
sdn-sc-04 & 2x E5-2667 v4 @ 3.20GHz & 16 & 128 & 100 \tabularnewline \hline
sdn-sc-05 & 2x E5-2667 v4 @ 3.20GHz & 16 & 128 & 100 \tabularnewline \hline
sdn-sc-06 & 2x E5-2667 v4 @ 3.20GHz & 16 & 128 & 100 \tabularnewline \hline
sandie-9 & 2x E5-2667 v3 @ 3.20GHz & 16 & 128 & 100 \tabularnewline \hline
\end{tabular}
\label{tb:specs-caltec}
\end{table}

\begin{table}[htb!] 
\vspace{-0.2cm} 
\scriptsize
\centering
\caption{UCSD servers specifications}
\vspace{-0.3cm} 
\begin{tabular}{|r|r|r|r|r|} \hline
{\shortstack{Name}} & {\shortstack{CPU model}} &  {\shortstack{\# of real\\cores}} & {\shortstack{RAM\\TB}} & {\shortstack{Total bandwidth\\capacity Gbps}} \tabularnewline \hline \hline
k8s-gen4-01 & 2x AMD EPYC 7763 @ 2.4GHz  & 128 & 2 & 500 \tabularnewline \hline
k8s-gen4-02 & 2x AMD EPYC 7763 @ 2.4GHz  & 128 & 2 & 700 \tabularnewline \hline
\end{tabular}
\label{tb:specs-ucsd}
\end{table}

All servers are deployed and managed using Kubernetes and have a non-shared, tmpfs file system mounted using the Kubernetes emptyDir volume directive with the medium defined to be Memory \cite{KubernetesStorageVolumes}. This helps avoid a likely bottleneck imposed by a slow file system. 
The test files are 1~GB each and are created using the Unix \textit{dd} command prior to starting the XRootD service.

\subsection{XRootD Configuration} 
\label{subsec:xrd_config}
The XRootD servers are configured to mimic the CMS production systems at UCSD and Caltech.
The servers are configured with the XRootD http directive and use a combination of X509 and Macaroon tokens for authentication/authorization.
X509 is used initially by gfal-copy~\cite{gfal} to obtain tokens from both source and destination sites, then the TPC is performed using only Macaroon tokens. This follows the current model used by CMS production systems.

\subsection{Transfer Orchestrator} 
\label{subsec:transfers}
The transfers are orchestrated using a bash script running third-party-copies using gfal-copy. This script, although simplistic in design, drains a significant amount of CPU and thus has to be run on a separate server. The transfers run concurrently, with each gfal-copy client running on a separate thread launching single stream transfers. The transfers are run between the XRootD clusters using the Redirector as the primary endpoint.

\section{Results} 
\label{sec:result}

\subsection{Achieving 400 Gbps} 
\label{subsec:400}

\begin{figure}[h!]
    \centering
    \includegraphics[width=\textwidth]{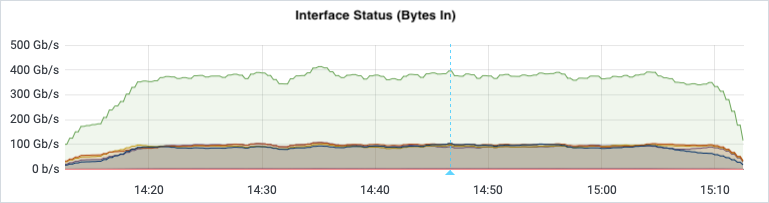}
    \includegraphics[width=\textwidth]{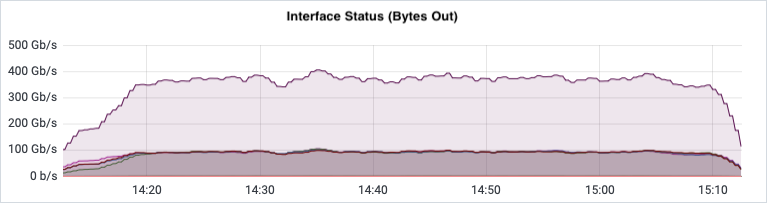}
    \caption{Monitoring plots showing 400~Gbps of throughput through the Arista switch. The plot on top shows traffic flowing into the Arista and the bottom plot shows throughput out of the port channel towards Caltech. The throughput through each separate port is showcased in the lines peaking at 100~Gbps and the line at 400~Gbps shows the aggregate.}
    \label{fig:400-plot}
\end{figure}

By virtue of trial and error, we determined that 40 streams per destination server (i.e. 40 $\times$ 13 = 520 streams coming out of UCSD) were required to saturate the bandwidth of the network link between UCSD and Caltech. Configuring our TPC orchestrator to produce this number of active transfers continuously we were able to generate and maintain an aggregated throughput of 400 Gpbs flowing through our testbed. Figure~\ref{fig:400-plot} shows the network traffic going out of the PortChannel towards Caltech on the Arista switch (marked in red in Figure~\ref{fig:400-diagram}). The throughput was sustained for around 1 hour and overall looks stable with little fluctuations.

\subsection{Extended Study}
\label{subsec:extended}
We try to go a step further and parameterize the throughput we achieve between two endpoints by the latency between the endpoints, number of streams, number of CPU cores used by each server, and the number of XRootD origins in use. This will, in a perfect world, give us an equation for the minimum values of each of these parameters required to attain a specific throughput.

In order to do this, we systematically vary these parameters and run transfers between two identical hosts at UCSD (k8s-gen4-01 and k8s-gen4-02 as listed in Table~\ref{tb:specs-ucsd}). The transfers are run until the throughput stabilizes and then are killed. We use the raw latency between the hosts (which is 0.1ms) as the baseline, and simulate different latencies using Linux traffic control (tc) to introduce artificial delay by holding packets in a buffer for a specified amount of time before allowing them to be transmitted. The number of CPU cores in the cluster are controlled using Kubernetes resource allocation.


We observed that a machine reboot was required in order to completely
reverse the network delays introduced through the use of the tc command.

\begin{figure}[h]
    \centering
    \includegraphics[width=\textwidth]{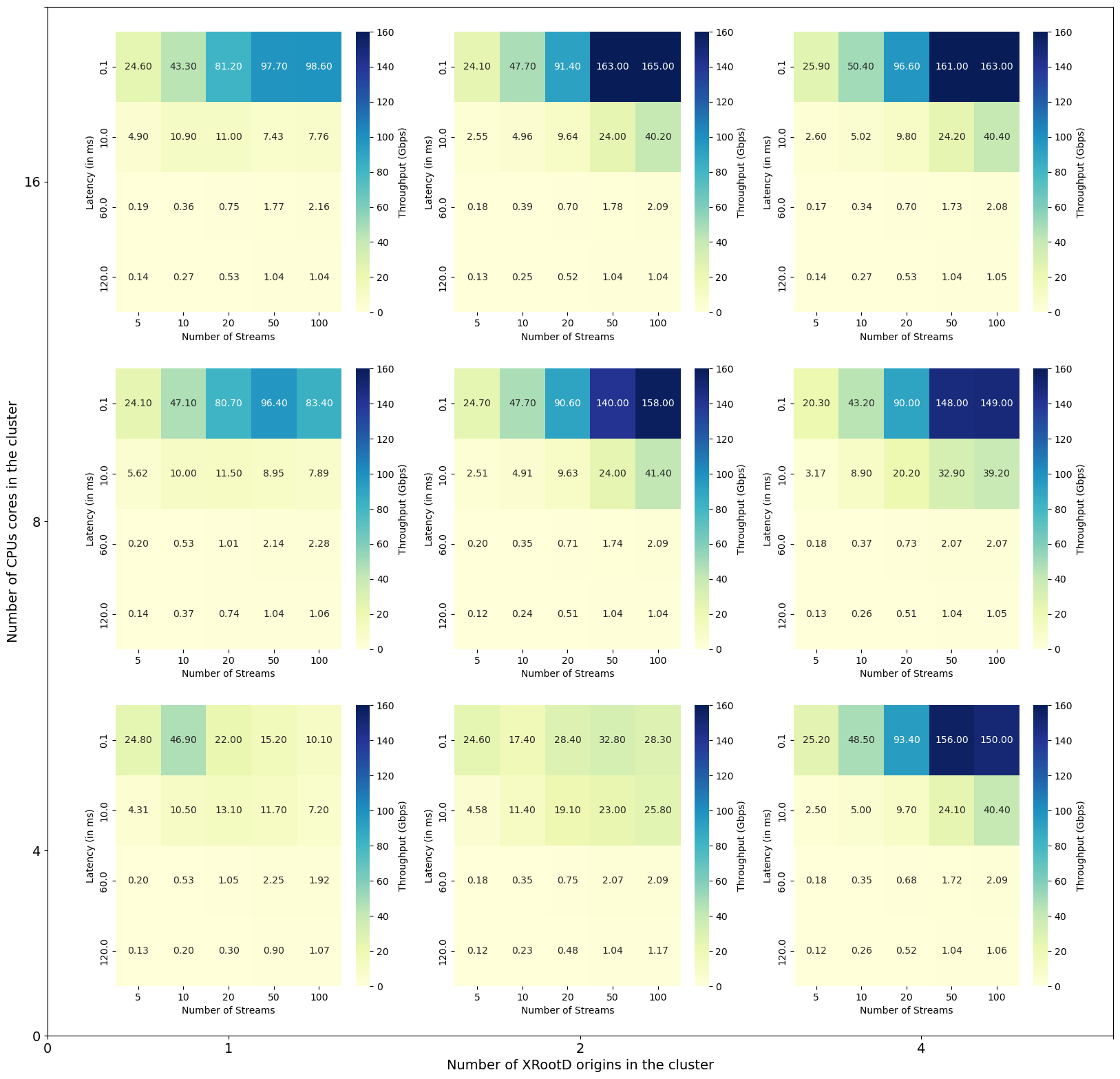}
    \caption{Plots showing Throughput (in Gbps) as a function of Number of streams and Latency (in ms) for different combinations of CPU cores and XRootD origins}
    \label{fig:latency}
\end{figure}

Each server uses its own network interface card (NIC) which has a bandwidth capacity of 100 Gbps, therefore increasing the number of XRootD origins increases the total bandwidth capacity accordingly. The number of CPU cores are aggregated over all origins in the cluster. The total number of streams are distributed across the origins, i.e. 100 streams with 4 servers would mean each server transferring using 25 streams. The results are showcased in the plots in Figure~\ref{fig:latency}

\subsection{More Lessons Learned}
\label{subsec:other}

\subsubsection{Overhead due to using Redirector}
Using the Redirector for load balancing as opposed to point to point transfers between the origins, even for large number of concurrent transfers causes no major overhead. Performance difference between the scenario where the Redirector manages which origin the request goes to, as opposed to when the source and destination server pairs are chosen by hand is negligible thus making it evident that using the Redirector is feasible.

\subsubsection{Choice of client transfer tool}
The choice of client transfer tool does make a slight difference. When testing the TPCs using gfal-copy and curl, we see $15\%$ better performance using gfal-copy. Thus other production tools like FTS \cite{fts} that employ the gfal client for TPCs should give performance similar to what we see.

\section{Conclusion} 
\label{sec:conclusion}
\subsection{Extended Study Results Interpretation}
As was observed in our previous study \cite{Fajardo2021}, the relationship between throughput and number of streams holds. As can be seen in all the plots in Figure~\ref{fig:latency}, for a low number of CPU cores, the throughput initially increases with the number of streams and then falls when the CPU usage is very high. This is likely due to resource contention or slow packet processing. Therefore, there is a maximum throughput that can be achieved using a given number of CPU cores.

Next we see that, adding more origins to the cluster even while keeping the number of cores the same (i.e. traversing in the rightward direction in any row in the plot) provides better performance which might owe to better kernel load balancing due to a higher number of processes.

We also conclusively see that throughput decreases severely with increasing latency. Going from 0.1 to 10 ms causes a five fold decrease in the amount of data transferred. It is therefore evident, as we would expect, that in order to achieve high throughput at very high latency, a very large number of streams would be required. We must note, however, that we do not see the theoretical linear scaling of throughput with latency whilst keeping the number of streams constant. This might point to the fact that tc is not ideal for delay emulation in high bandwidth tests. This suspicion is further deepened by the fact that repeating the transfer test with two origins, each having access to all 128 cores on the host and pushing up to 500 streams, the throughput at 120ms of latency is still restricted to around 1~Gbps.

\subsection{Overall Conclusion}

The results show that XRootD HTTPS-TPC is capable of sustaining the high throughput required in the High Luminosity LHC era at low latencies. The trends in our study seem to indicate that with the right combination of streams, number of cpu cores and number of origins in the cluster, the desired throughput can be achieved for higher latencies, but more work is required to be completely certain.

\section{Future Work}
More work is required to understand the deviation from the expected trend when we increase the latency from 10ms to 60ms and also why throughput scaling with the number of streams stops at higher latencies.

\section{Acknowledgement} 
\label{sec:acknowledgement}
The authors would like to thank CENIC for providing us with the network bandwidth that made this work possible, ARISTA for donating to us the switch  in CENIC LA 818, 10th floor, and the funding agencies to support this effort, in particular the National Science Foundation through the following grants: OAC-1836650, PHY-2323298, OAC-2030508 and OAC-2112167

\bibliography{references.bib}

\end{document}